\documentclass{cernrep}

\usepackage{xcolor}

\begin{document}
\title{Exclusive meets inclusive at small Bjorken-$x_B$: how to relate exclusive measurements to PDFs based on evolution equations }
\author{Hervé Dutrieux, Michael Winn and Valerio Bertone}
\institute{IRFU, CEA, Université Paris-Saclay, 91191 Gif-sur-Yvette, France}

\begin{abstract}
Exclusive heavy-vector-meson photoproduction is a prominent signal in collider experiments with hadron beams. At the highest photon-hadron collision energies, this process is considered as a candidate to constrain gluon parton distribution functions~(PDFs) at small longitudinal momentum fractions. However, in the framework of collinear factorisation, exclusive particle production is described in terms of generalised parton distributions~(GPDs).  In this contribution, we investigate at the leading order in $\alpha_s$ the connection between GPDs and PDFs. Our main result is a proposal to quantify the systematic uncertainty inherent to this connection. We put our approach into context with respect to the Shuavev transform. Our uncertainty estimate can be straightforwardly adapted to higher fixed orders and small-$x$ resummations. The question of extrapolating GPDs to vanishing skewness is paramount for the programme of the Electron Ion Collider (EIC), notably for the extraction of the radial distributions of partons.

\end{abstract}


\maketitle

\section{Introduction}

Partons at very small longitudinal momentum fractions and high energies constitute a particularly interesting regime of quantum chromodynamics (QCD), where so-called gluon saturation has been predicted to occur~\cite{Gelis:2010nm}: hadrons are no longer a dilute collections of partons, but interacting systems dominated by gluons. Calculations of event rates in ultra-high-energy neutrino astrophysics also depend on hadron and nuclear structure at very small longitudinal momentum fraction~\cite{Cooper-Sarkar:2011jtt,Bertone:2018dse}. Additionally, partons in this regime define the initial state of the thermodynamic system created in heavy-ion and in general hadron collisions at colliders~\cite{Gelis:2010nm}. It is known from experiment that gluons dominate the partonic content at small-$x$~\cite{Ethier:2020way, Gao:2017yyd}. However, the uncertainty on the gluon parton distribution function (PDF) for longitudinal momentum fractions $x \sim 10^{-4}$ or less is still significant, mainly because experimental access is scarcer than for larger values of $x$. Experimental sensitivity can be achieved through different measurements at collider facilities which provide the largest collision energies $\sqrt{s}$. For instance, at the LHC inclusive particle production, in particular of charm and beauty quarks, has been proposed to constrain gluon PDF in the proton~\cite{Gauld:2015yia,Gauld:2015kvh,Garzelli:2016xmx,Zenaiev:2019ktw,Khalek:2022zqe}. Although a noticeable reduction of uncertainty is produced by the inclusion of these measurements, their impact is limited by missing higher-order corrections as indicated by the large scale and hadronization uncertainties, and other effects not accounted for in perturbative QCD (pQCD). In addition to these observables, exclusive hard photoproduction processes which can also be measured at hadron colliders are interesting to consider since they are expected to be less, or at least, differently, affected by phenomena not accounted for in state-of-the-art pQCD calculations. 

Of particular interest is exclusive heavy-vector-meson production (HVMP) of $J/\psi$ or $\Upsilon$ mesons, where the hard scale is provided by the mass of the vector meson $m_{V}$ itself, and typical values of Bjorken's $x_B$\footnote{We stress the difference between the parton longitudinal momentum fraction $x$ on which parton distributions depend, and Bjorken's $x_B$ which is a kinematic variable at which an observable is measured. Many observables measured at $x_B$ exhibit a strong sensitivity to PDFs and GPDs at values $x$ of the order of $x_B$, hence small-$x$ and small-$x_B$ are often used interchangeably. However, in general, factorised observables depend on an integral of parton distributions over a range of values of $x$, typically up to $x = 1$.} are of the order of $e^{-y} m_V / \sqrt{s}$ where $y$ is the pseudorapidity of the meson. HERA~\cite{ZEUS:1998fem,ZEUS:2002wfj,H1:2002yab,Chekanov_2004,H1:2005dtp,ZEUS:2009ixm,H1:2013okq}, the LHC~\cite{Aaij:2013jxj,Aaij:2014iea,TheALICE:2014dwa,Aaij:2015kea,Aaij:2018arx,Sirunyan:2018sav} or a future LHeC allow for measurements at small values of Bjorken's $x_B$, of the order of $10^{-4}$ down to $10^{-6}$. The future EIC~\cite{AbdulKhalek:2021gbh} will provide precise HVMP data with Bjorken's $x_B$ of the order of $10^{-3}$ to $10^{-4}$. 

\begin{figure}
    \centering
    \includegraphics[scale=0.17]{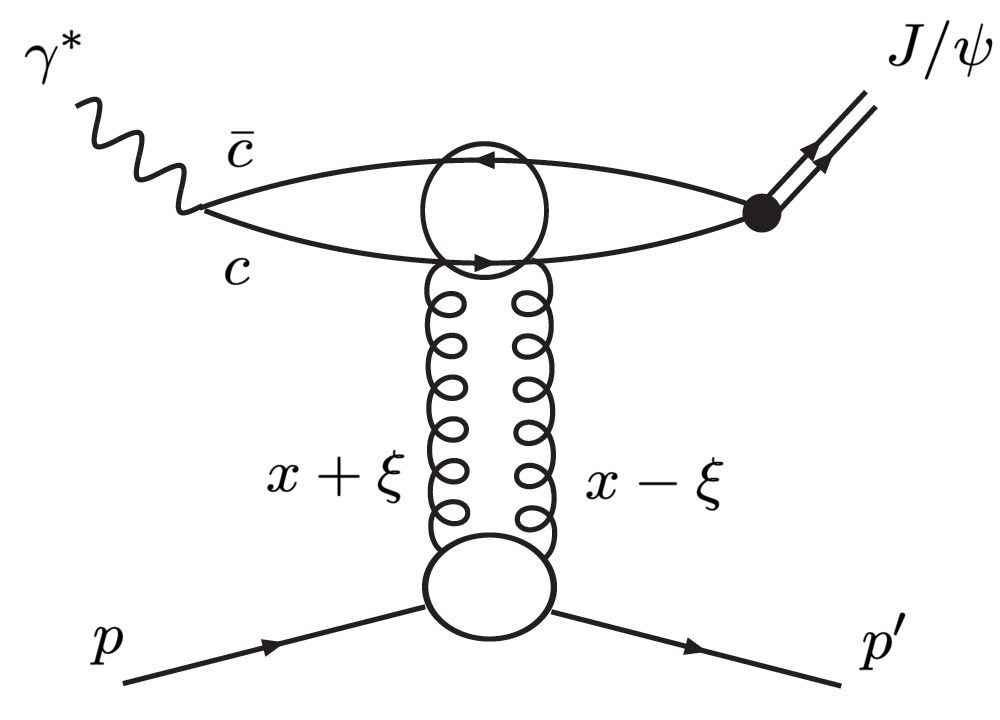}
    \caption{Photoproduction of a heavy vector meson $J/\psi$ (adapted from Ref.~\cite{Jones:2013pga}).}
    \label{fig:photoproduction}
\end{figure}

The leading-order (LO) two-gluon exchange in the $t$-channel depicted in Fig.~\ref{fig:photoproduction} is the dominant contribution to the HVMP cross section. However, because there is a transfer of four-momentum between initial and final-state hadrons, $p$ and $p'$, the description of this process in the framework of collinear factorization~\cite{Ivanov:2004vd} does not involve usual PDFs, but the so-called generalized parton distributions (GPDs)~\cite{Mueller:1998fv, Ji:1996ek, Ji:1996nm, Radyushkin:1996ru, Radyushkin:1997ki}. It is therefore desirable to establish a reconstruction procedure of GPDs in the small-$x$ region from PDFs in order to extract information on PDFs from HVMP data. The study of this procedure has attracted interest since the early days of GPD studies, more than two decades ago, and is the main objective of this paper. One of the major contributions, based on the Shuvaev transform, was proposed in Refs.~\cite{Shuvaev:1999fm,Shuvaev:1999ce} and was applied to HERA and LHC experimental data for instance in Refs.~\cite{Jones:2013pga, Flett:2019pux, Flett:2021fvo}. In this article, we suggest an alternative procedure that allows us to evaluate the theoretical uncertainty associated to linking GPDs to PDFs and to pave the way for more detailed studies at higher orders and with small-$x$ resummation.

The outline of this article is as follows. We start by exposing the problem of relating GPDs to PDFs and we explicit the role that evolution equations play in that respect. We also briefly highlight the relevance of this issue for the physics programme of the EIC, independently of the question of HVMP. Next, we present our proposal to quantify the uncertainty on relating GPDs to PDFs, and derive an estimate of the uncertainty for $J/\psi$ and $\Upsilon$ HVMP. In the last section, we finally connect our method to the Shuvaev transform.

\section{Kinematic variables, evolution operators, and the link between GPDs and PDFs}

We only sketch out the properties of GPDs necessary to provide a sufficient context.  More details can be found for instance in the review articles in Refs.~\cite{Diehl:2003ny, Belitsky:2005qn, Boffi:2007yc}. We use the notations of Ref.~\cite{Diehl:2003ny} for the gluon GPD $F^g$. A GPD $F^a$ for a given parton of type $a = q, g, \hdots$ in a hadron is a function of the average longitudinal momentum fraction of the parton itself denoted by $x$, the Mandelstam variable $t = (p'-p)^2$, where $p$ and $p'$ are the incoming and outgoing hadron four-momenta respectively, and the skewness $\xi$, which measures the longitudinal-momentum transfer and can be related to Bjorken's $x_B$ in the Bjorken limit through
\begin{equation}
    x_B \approx \frac{2\xi}{1+\xi}\,.
\end{equation}
In the following, we will use the skewness variable $\xi$ as is customary in GPD studies, keeping in mind that in the regime of small Bjorken's $x_B$, $\xi \approx x_B / 2$. The regime of small-$x_B$ is therefore the regime of small-$\xi$ in GPD terminology. 

In addition to $(x, \xi, t)$, GPDs also depend on a factorization scale denoted by $\mu$. The relation between the GPD $F^a$ and the corresponding PDF $f^a$ for the same parton and hadron helicities in initial and final states is given by
\begin{align}
    F^a(x, \xi = 0, t = 0, \mu) &= x^{p_a}f^a(x, \mu) \,,
    \label{eq:forwardlimit}
\end{align}
where $p_a = 0$ for quarks $(a = q)$ and 1 for gluons $(a = g)$. As discussed in the previous section, our main objective is to reconstruct the GPDs at small-$x$ and $x_B$ (that is small-$x$ and $\xi$) from PDFs.  Eq.~\eqref{eq:forwardlimit} gives a straightforward answer even if $\xi$ is not strictly zero, but $x \gg \xi$ and $t$ is negligible compared to the hard scale of the process. Then the transfer of four-momentum may be neglected altogether, and the GPD approximated by the PDF. If $x \gg \xi$ but $t$ is not negligible, relating GPDs to PDFs requires control over the $t$ behaviour, usually expressed by an exponential Ansatz for the $t$-dependence as for instance in Ref.~\cite{Jones:2013pga}. A study of the variation of the $t$-dependence in some models under evolution can be found in Ref.~\cite{Diehl:2007zu}. 

Since the uncertainty on the relation between GPDs and PDFs at non-vanishing $t$ can be adequately addressed by flexible parametrisations of the $t$-dependence, we will set this aspect aside and focus on linking GPDs and PDFs when $\xi$ is non-negligible as compared to $x$. This case deserves a particular attention for several reasons. 

\begin{enumerate}
    \item Evolution equations produce a direct entanglement of the $x$, $\xi$, and $\mu$ dependences of GPDs, whereas the evolution kernels are independent of $t$. 
    \item The convolutions of GPDs that enter the description of exclusive HVMP exhibit a strong sensitivity to the region $x \approx \xi$. Therefore, the region where GPDs cannot be bluntly approximated by PDFs is precisely the one we are mostly concerned with.
    \item GPDs with vanishing skewness $F^a(x, \xi = 0, t)$ are of a great importance, as they are used for the definition of impact parameter distributions (IPDs)~\cite{Burkardt:2000za} giving the probability density of a parton carrying a fraction of longitudinal momentum $x$ at some radial distance $b_\perp$ from the center of momentum of the hadron in the infinite momentum frame:
    \begin{equation}
        \mathrm{IPD}^a(x, b_\perp, \mu) = \int \frac{\mathrm{d}^2 \Delta_\perp}{(2\pi)^2}\,e^{-ib_\perp\cdot\Delta_\perp} F^a\left(x, \xi = 0, t = -\Delta_\perp^2,\mu\right)\,.
        \label{eq:ipds}
    \end{equation}
    The imaging of the hadronic structure in position space has been identified as one of the goals of the EIC programme~\cite{AbdulKhalek:2021gbh}. As most data on exclusive processes are taken at non-vanishing skewness, an understanding of the uncertainty associated to the extrapolation to $\xi = 0$ at fixed value of $t$ is crucial to this purpose, and is provided by the procedure we present in this article.
\end{enumerate}

Since HVMP is only sensitive to $C$-even GPDs, in the following we will consider only such GPDs, also known as singlet GPDs. They have a definite parity in $x$ - gluon GPDs are $x$-even and quark singlet GPDs are $x$-odd - which allows us to only focus on the case where $x \geq 0$. In general, GPDs are also even functions of $\xi$ due to time reversal invariance, so we consider only $\xi \geq 0$ as well.

The dependence of PDFs on their scale $\mu$ is given by the Dokshitzer-Gribov-Lipatov-Altarelli-Parisi (DGLAP) renormalization group equations~\cite{Dokshitzer:1977sg, Gribov:1972ri, Altarelli:1977zs} which ensure that observables computed from PDFs do not depend on the arbitrary choice of $\mu$. GPDs follow their own evolution equations~\cite{Mueller:1998fv, Ji:1996nm, Radyushkin:1997ki}, which generalize the DGLAP equations:\footnote{Following Eq.~\eqref{eq:forwardlimit}, $x^{-p_a}F^a(x, \xi, t, \mu)$ where $p_a = 1$ if $a = g$ and 0 otherwise gives exactly the correct forward limit $f^a(x,\mu)$ whether $a = q$ or $g$.}
\begin{align}
    \frac{F^a(x, \xi, t, \mu)}{x^{p_a}} &= \sum_{b = q, g, ...}\int_0^1 \frac{\mathrm{d}z}{x}\,\Gamma^{ab}\left(\frac{z}{x}, \frac{\xi}{x}; \mu_0, \mu\right) \frac{F^b(z, \xi, t, \mu_0)}{z^{p_b}}\,.
    \label{eq:opdef}
\end{align} 
In light of the definition in Eq.~\eqref{eq:opdef}, the usual DGLAP integrated evolution equations can be written as
\begin{align}
    f^a(x, \mu) &= \sum_{b = q, g, ...}\int_0^1 \frac{\mathrm{d}z}{x}\,\Gamma_0^{ab}\left(\frac{z}{x}; \mu_0, \mu\right) f^b(z, \mu_0)\,,
    \label{eq:oppdf}
\end{align}
where the DGLAP evolution operator $\Gamma_0^{ab}$ is linked to the general one $\Gamma^{ab}$ through
\begin{equation}
    \Gamma_0^{ab}\left(\frac{z}{x}; \mu_0, \mu\right) = \lim_{\xi/x \rightarrow 0} \Gamma^{ab}\left(\frac{z}{x}, \frac{\xi}{x}; \mu_0, \mu\right)\,.
\end{equation}
The DGLAP operator $\Gamma_0^{ab}(z/x; \mu_0, \mu)$ takes non-zero values for $\mu \geq \mu_0$ if and only if $z \geq x$. Intuitively, this can be understood as follows: since the DGLAP evolution describes the substructure of partons as the resolution at which the system is probed increases, a parton carrying a longitudinal-momentum fraction $x$ at the initial scale radiates several partons ending up with a momentum fraction less than $x$ at the final scale. In practice, as we will show below, DGLAP evolution, and the general GPD evolution for $x \geq \xi$ as well, pushes distributions from large- to small-$x$. As a result, the higher the evolution range is, the more variations of the distribution at initial scale are washed away towards a smaller-$x$ domain. This general idea forms the basis of the strategy to reconstruct the $\xi$-dependence of GPDs from PDFs: \textit{i.e.} we ignore the $\xi$-dependence of the GPD at an initial low-lying scale in that the evolution to sufficiently higher scales washes away this dependence.

\begin{figure}
    \centering
    \includegraphics[scale=0.7]{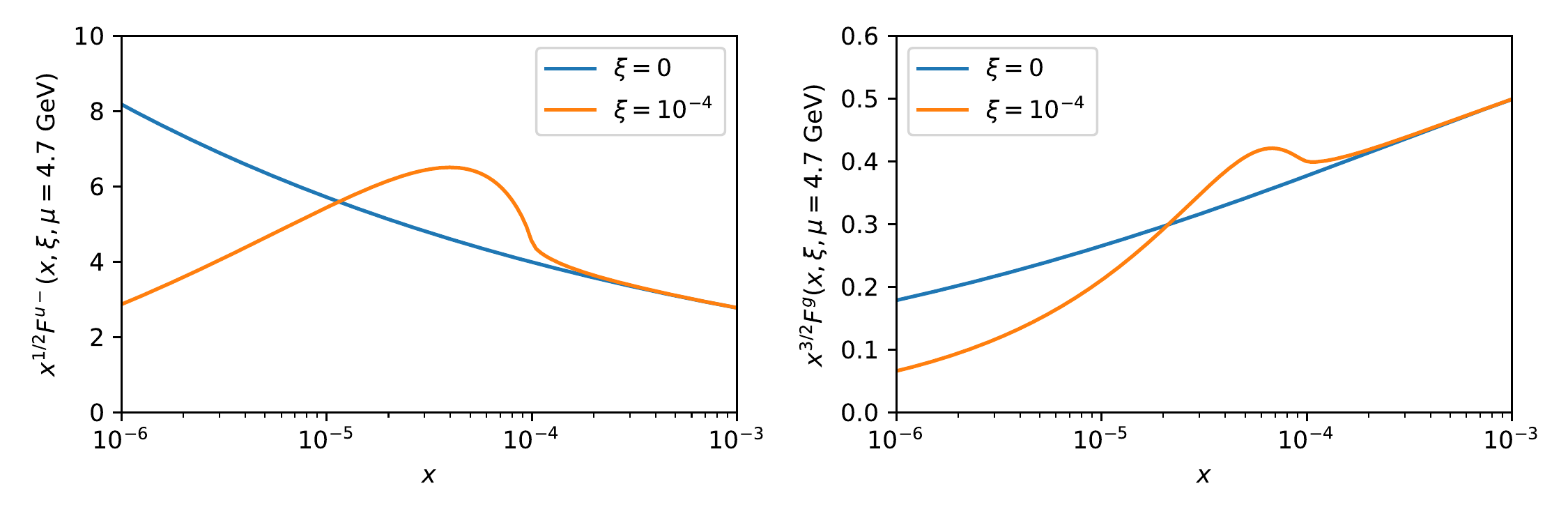}
    \caption{Evolution of the MMHT2014 LO PDF~\cite{Harland-Lang:2014zoa} from $\mu_0 = 1$~GeV to 4.7~GeV using the DGLAP evolution (blue curve) and the GPD evolution at $\xi = 10^{-4}$ (orange curve). On the left, we show the non-singlet $u$ PDF defined by $u^-(x) = u(x) - \bar{u}(x)$. On the right, the gluon PDF. The difference between the two curves becomes sizeable for $x \lesssim 3\xi$.}
    \label{fig:pdfevol}
\end{figure}

Before undertaking a discussion of the $\xi$-dependence of the LO evolution operator, it is enlightening to observe directly its consequences. In Fig.~\ref{fig:pdfevol}, we use the LO PDF set MMHT2014~\cite{Harland-Lang:2014zoa} as an input at the scale $\mu_0 = 1$~GeV, and evolve it to $\mu = 4.7$~GeV using both the ordinary LO DGLAP equation ($\xi = 0$), and the LO GPD evolution equation at $\xi = 10^{-4}$. In accordance with the MMHT2014 LO extraction, the running of the strong coupling is computed at one loop with $\alpha_s(M_Z) = 0.135$ in the variable-flavour-number scheme, with the charm threshold at 1.4~GeV. As we use the same distribution at initial scale, the entire difference between the curves in Fig.~\ref{fig:pdfevol} is due to the $\xi$-dependence introduced by the evolution operator. One can observe immediately that the difference is only perceptible for $x \lesssim 3\xi$. As we have already mentioned, this is the region of largest phenomenological interest for the description of HVMP data. To rephrase the strategy presented at the end of the previous paragraph, evolution is expected to wash away the uncertainty due to the unknown (non-perturbative) $\xi$-dependence of the GPD at initial scale, replacing it with the known (perturbative) $\xi$-dependence produced by the evolution operator.

\begin{figure}
    \centering
    \includegraphics[scale=0.7]{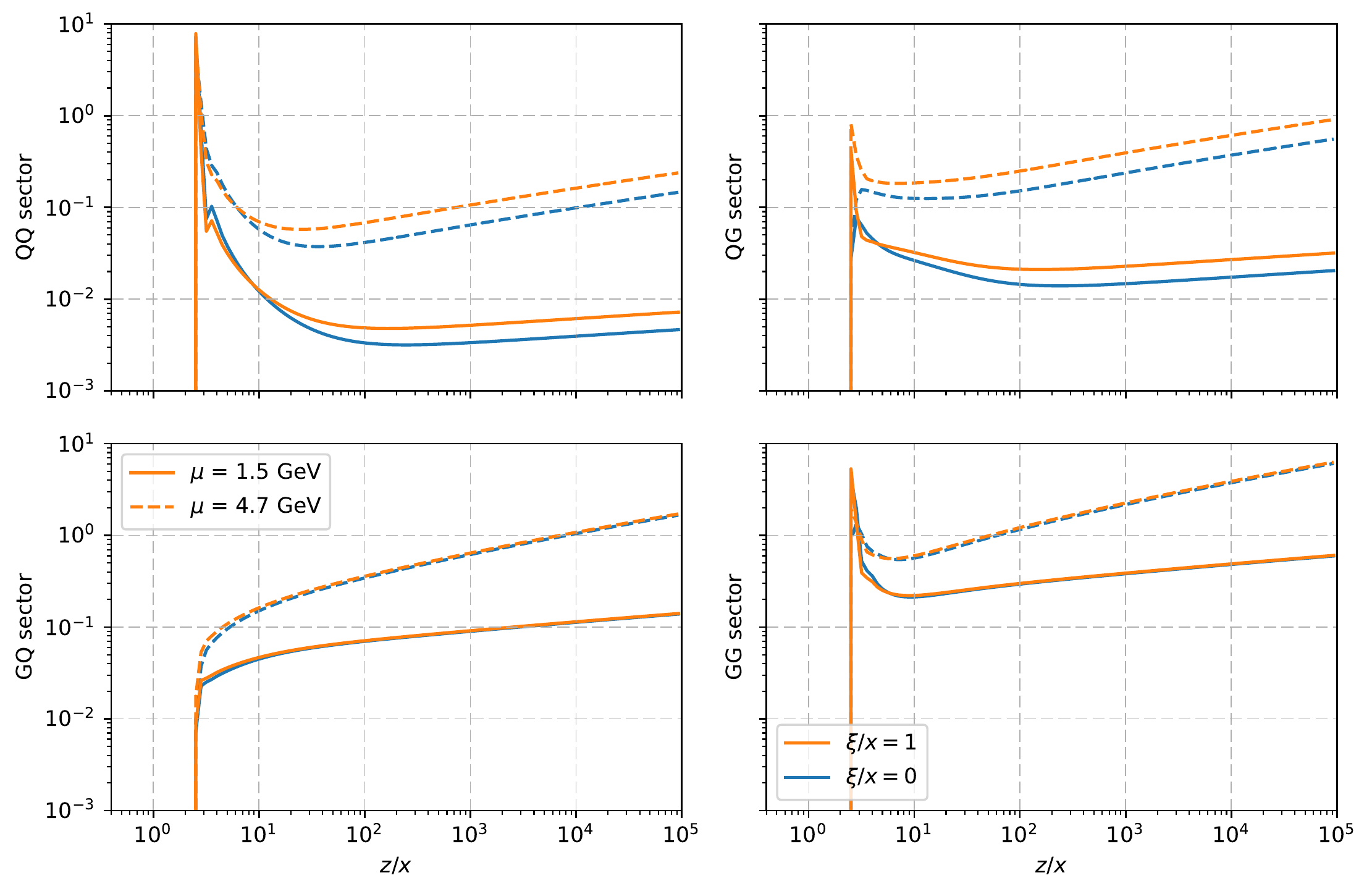}
    \caption{Behaviour of $\Gamma^{ab}(z/x, \xi / x;\mu_0$ = 1~GeV, $\mu)$ in the four sectors $ab = qq$ (upper left), $ab = qg$ (upper right), $ab = gq$ (lower left) and $ab = gg$ (lower left). The continuous lines stand for the final scale $\mu = m_{J/\psi}/2$ = 1.5~GeV while the dotted lines for $\mu = m_{\Upsilon}/2$ = 4.7 GeV. The colors refer to the value of the ratio $\xi / x$: $\xi = 0$ is the DGLAP operator (blue) and $x = \xi$ is in orange.}
    \label{fig:operators}
\end{figure}

Referring to Eq.~\eqref{eq:opdef}, the $\Gamma^{ab}$ operators give a weighting\footnote{We use evolution equations with non-infinitesimal distance between $\mu_0$ and $\mu$. It is more frequent to see them obtained by applying $\mathrm{d} / \mathrm{d} \log \mu$ to Eq.~\eqref{eq:opdef}. The differential form involves the derivative of $\Gamma^{ab}$ with respect to $\mu$, known as \textit{splitting function}, which is a distribution involving Dirac $\delta$-functions and plus-prescriptions. Using integrated evolution equations allows one to treat $\Gamma^{ab}$ as an ordinary weighting function, and not as a mathematical distribution (if $\mu \neq \mu_0$).} of GPDs at initial scale $\mu_0$ to produce GPDs at final scale $\mu$. As such, they allow one to gauge the importance of the contribution to the evolution of various regions of the GPD at the initial scale. The properties of evolution at small values of $\xi$ have notably been studied in Refs.~\cite{Frankfurt:1997ha, Martin:1997wy, Shuvaev:1999ce, Diehl:2007zu}, but to the best of our knowledge, studies of this kind have been performed by means of models of GPDs and DDs, often with the assumption of power-law
behaviour at small-$x$, and have therefore a lesser generality than our discussion which
takes place directly at the level of the $\Gamma^{ab}$ operators. We use the recent GPD evolution software APFEL++~\cite{Bertone:2013vaa, Bertone:2017gds, Bertone:2022frx} to study numerically their properties. For numerical applications, we will use the typical hard scales encountered in HVMP, given by half the mass of the vector meson~\cite{Jones:2015nna, Jones:2016ldq}, that is $\mu_c = m_{J/\psi} / 2 = 1.5$~GeV and $\mu_b = m_{\Upsilon} / 2 = 4.7$~GeV. Since we only perform LO evolution, we need an initial scale to start evolution which is high enough to produce coherent results. We choose as an initial scale $\mu_0 = 1$~GeV and plot in Fig.~\ref{fig:operators} the evolution operators $\Gamma^{ab}(z/x, \xi / x; \mu_0, \mu)$ as functions of $z/x$ evaluated at $\mu = \mu_c$ and $\mu_b$, with the more phenomenologically accurate $\alpha_s(M_Z) = 0.118$ and still a variable-flavour-number scheme. Let us comment on some important features.

\begin{enumerate}
\item We only focus on the case where $\xi \leq x$.\footnote{This region is sometimes called the DGLAP region, due precisely to the fact that the GPD evolution equation for $\xi \leq x$ shares many similarities with the DGLAP evolution equation. On the contrary, the region where $\xi \geq x$ is called ERBL thanks to a parallel with the ERBL evolution equation of distribution amplitudes~\cite{Efremov:1979qk, Lepage:1979zb}. We don't use these appellations to reserve the term DGLAP to the sole evolution equations of PDFs.} A specific procedure, called covariant extension~\cite{Chouika:2017dhe, Chouika:2017rzs}, allows one to retrieve the $\xi > x$ region from the $\xi \leq x$ one, up to a specific function known as the $D$-term. Since the $D$-term only lives in the region $\xi > x$ and enjoys its own independent evolution equation from the rest of the GPD, it does not provide any handle on the PDF. Therefore, the $D$-term contribution to the amplitude of a process, which can be isolated in the formalism of dispersion relations (see for instance Refs.~\cite{Anikin:2007yh, Diehl:2007jb}), is ignored in the present study. The region $\xi \leq x$ possesses another important property: $F^a(x, \xi, t, \mu)$ only depends on values of $F^{b}(z, \xi, t, \mu_0)$ such that $z \geq x$, as can be observed in Fig.~\ref{fig:operators}. We have already mentioned this property in the case of the DGLAP operator ($\xi = 0$).

\item The $qq$ and $gg$ sectors exhibit a strong peak at $z = x$, all the more that $\mu$ is close to $\mu_0$. This is easily understandable because in the limit of no evolution, where $\mu = \mu_0$, $\Gamma^{ab}(z/x, \xi/x;\mu_0, \mu_0) = \delta_{a,b} \delta(1-z/x)$ where $\delta_{a,b}$ is the Kronecker delta. On the contrary, as $\mu$ increases, the contribution from the region $z \gg x$ becomes increasingly important. This corresponds to a kinematic region where at the initial scale $\mu_0$, the asymmetry between the four-momenta of incoming and outgoing hadrons, $z + \xi$ and $z - \xi$, is negligible and the GPD can safely be replaced by the PDF. This finally provides a quantitatively proof of the argument that the evolution washes away the uncertainty on the $\xi$ behaviour of the GPD at the initial scale. Indeed, as the behaviour of the evolved GPD is increasingly controlled by its large-$z$ region at the initial scale, the initial uncertainty on its $\xi$-dependence becomes negligible.

\item Finally, one can observe that the DGLAP operator ($\xi = 0$) and the GPD evolution operator at $x = \xi$ share globally a similar shape. Nonetheless, the curves in the $gg$ and $gq$ sectors, which seem almost identical for $z / x \gtrsim 5$, actually differ by an almost constant factor of 5\% between $\xi = x$ and $\xi = 0$ in the large $z / x$ domain at $\mu = 4.7$~GeV. The difference is much starker in the $qq$ and $qg$ sectors, with an almost constant factor of 50\% for the same parameters. Significant differences are furthermore observed in the region where $z$ is of the order of $x$. This explains why in Fig.~\ref{fig:pdfevol} the GPD evolution at $\xi = 10^{-4}$ produces larger values than the DGLAP evolution around $x = \xi$. That the region around $x = \xi$ would evolve faster than the PDF was already identified in a number of early references~\cite{Frankfurt:1997ha, Shuvaev:1999ce, Musatov:1999xp} and often taken into account in models with the help of an appropriately tuned "skewness ratio" $F^a(x, \xi = x) / F^a(x, \xi = 0)$.
\end{enumerate}

Let us now draw a few conclusions. From Figs.~\ref{fig:pdfevol} and~\ref{fig:operators}, it is clear that simply using the PDF instead of the GPD when $x \approx \xi$ brings potential GPD modelling uncertainties of the order of several tens of percents even when $\xi$ is arbitrarily small. The skewness ratios quoted in Refs.~\cite{Frankfurt:1997ha, Shuvaev:1999ce} reach in some cases even highers values where the GPD at $x = \xi$ is about 1.6 times as large as the corresponding PDF. A precise study of HVMP must deal with this source of uncertainty. A path to do so is already clear. At a very low-lying scales, the $\xi$-dependence of a GPD may be unknown, but as the GPD is evolved, the increasing dominance of the region $z \gg \xi$ allows for a reduction of the uncertainty. This handle on the uncertainty of the extrapolation to zero skewness will depend on the range in $\mu$, the value of $\xi$ and the $x$ profile of the GPD at the initial scale. With this in mind, we can lay down our proposal to relate GPDs to PDFs with a quantification of systematic uncertainties.

\section{Proposal to quantify the systematic uncertainty in relating GPDs to PDFs}

The model of reconstruction of the $\xi$-dependence of GPDs at small-$\xi$ that we suggest is straightforward: we simply propose to approximate the GPDs by the PDFs at the low scale $\mu_0$. Then we evolve them to the hard scale $\mu$ with the full $\xi$ dependent GPD evolution operator. Observables can then be computed with this object whose entire $\xi$-dependence has been generated by evolution. We will show in the next section that this model is conceptually close to the one based on the Shuvaev transform, although with notable differences.

\begin{figure}
    \centering
    \includegraphics[scale=0.65]{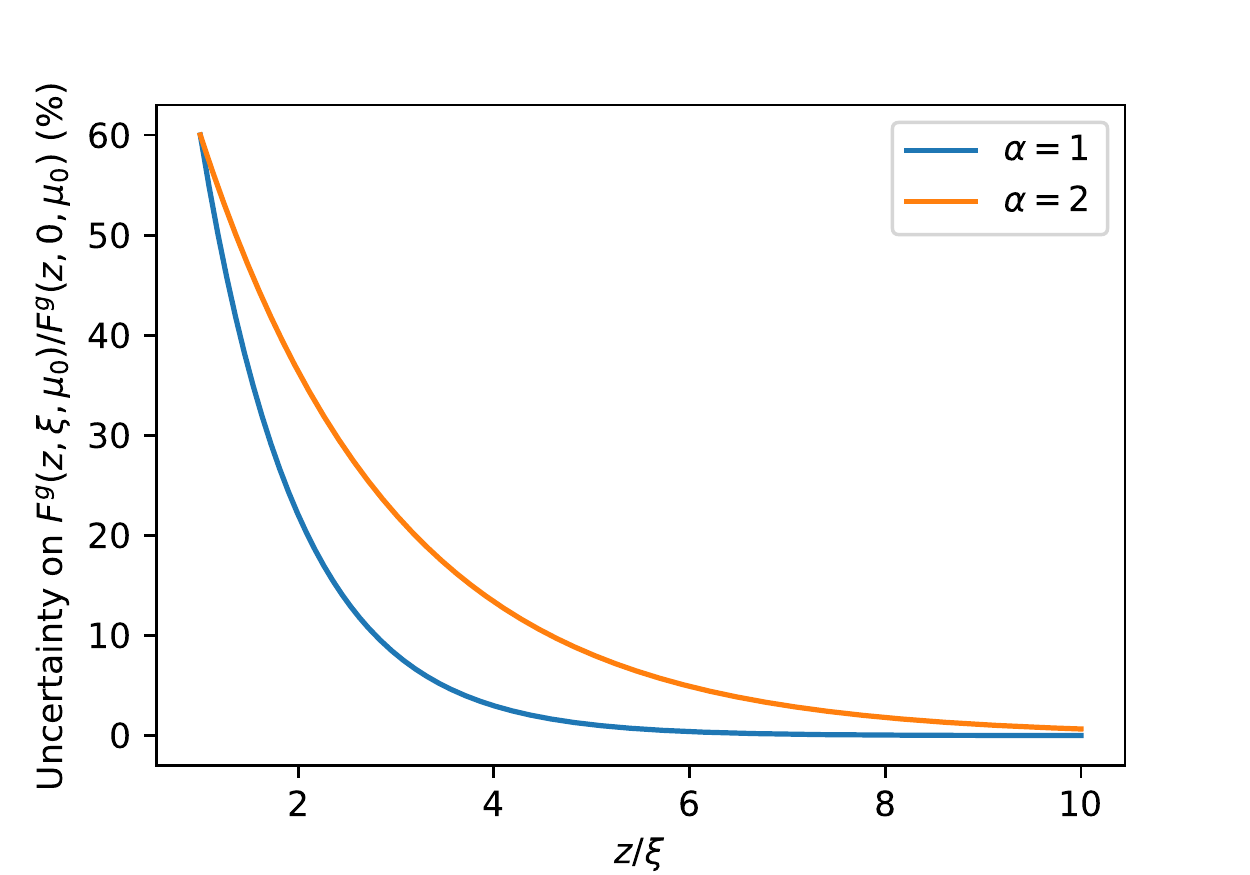}
    \caption{Uncertainty profiles produced by Eq.~\eqref{eq:unc_profile} with $\alpha=1$ (blue curve) and $\alpha=2$ (orange curve) to quantify the unknown $\xi$-dependence of the GPDs at the initial scale. In both cases, we assume that the GPD at $\xi = z$ deviates by 60\% from the corresponding PDF. For $z \gtrsim 10 \xi$, the difference between PDF and GPD is very small as one would expect.}
    \label{fig:unc}
\end{figure}

Beyond its simplicity, the main advantage of our proposal is the possibility to compute an estimate of the associated systematic uncertainty. Since we don't know the $\xi$-dependence of the GPDs at the initial scale, there is by definition a certain level of arbitrariness in any attempt to quantify the uncertainty of our procedure, which can be used as a conservative estimate of the magnitude of the uncertainty. At the initial scale $\mu_0$, pessimistic estimates derived from the skewness ratio evoked at the end of the previous section lead us to assume a plausible uncertainty of up to 60\% at $z = \xi$ between the GPD $F^a(z = \xi, \xi, \mu_0)$ and the PDF $f^a(z, \mu_0)$. We expect this uncertainty to decrease quickly as $z$ increases, and become essentially negligible for $z \gtrsim 10\xi$. We show in Fig.~\ref{fig:unc} two plausible profiles of uncertainty on the $\xi$-dependence of the GPD at the initial scale as a function of $z / \xi$, given by
\begin{equation}
    g\left(\frac{z}{\xi}\right) = 0.6 \exp\left(\frac{\xi-z}{\alpha\xi}\right)\,,
    \label{eq:unc_profile}
\end{equation}
with $\alpha = 1$ to simulate a fast decrease in uncertainty and $\alpha=2$ for a slower decrease. As we will see, this does not make much difference, indicating that it is the uncertainty at $z = \xi$ which represents the most critical source of uncertainty.

Then we assume $g(z/\xi) f^a(z, \mu_0)$ to be an upper bound on the uncertainty of the GPD at initial scale, and we can define our conservative magnitude of uncertainty at scale $\mu$ by simply integrating this uncertainty against the evolution operators, 
\begin{equation}
    \Delta^{a}(x, \xi, \mu) = \frac{\displaystyle \sum_{b} \int_{0}^1 \frac{\mathrm{d}z}{x}\, \Gamma^{ab}\left(\frac{z}{x}, \frac{\xi}{x};\mu_0, \mu\right)\,g\left(\frac{z}{\xi}\right)f^b(z, \mu_0)}{\displaystyle \sum_{b} \int_{0}^1 \frac{\mathrm{d}z}{x}\, \Gamma^{ab}\left(\frac{z}{x}, \frac{\xi}{x};\mu_0, \mu\right)\,f^b(z, \mu_0)}\,. \label{eq:unc_est}
\end{equation}
For instance, if $\mu = \mu_0$, which corresponds to no evolution, then the relevant evolution operators are just Dirac peaks centered at $x = z$, and 
\begin{equation}
    \Delta^a(x, \xi, \mu = \mu_0) =   \frac{\sum_{b} \delta_{a,b}g(x/\xi)f^b(x, \mu_0)}{\sum_{b} \delta_{a,b}f^b(x, \mu_0)} =  g\left(\frac{x}{\xi}\right)\,.\label{eq:fhoeizo}
\end{equation}
As expected, without evolution we recover exactly the uncertainty assumed at the initial scale given by the function $g$.

\begin{figure}
    \centering
    \includegraphics[scale=0.62]{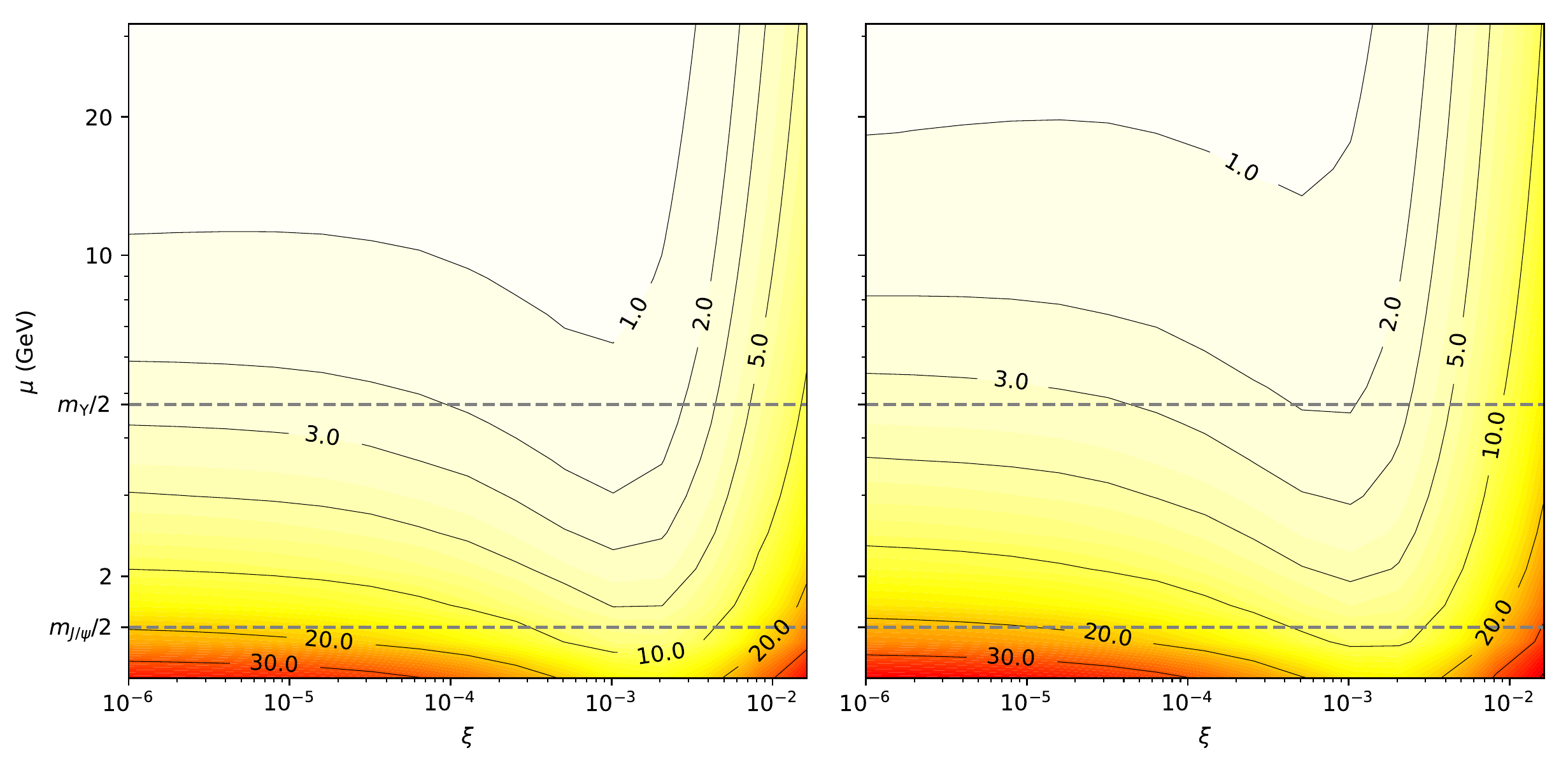}
    \caption{Behaviour of $\Delta^g(x = \xi, \xi, \mu)$ in \% as a function of $\xi$ and $\mu$ for the LO PDF set MMHT2014 used as input at the scale $\mu_0 = 1$~GeV. On the left, we use the uncertainty profile with $\alpha = 1$ while on the right with $\alpha = 2$ (see Fig.~\ref{fig:unc} and Eq.~\eqref{eq:unc_profile}).}
    \label{fig:prop_large_y}
\end{figure}

We display on Fig.~\ref{fig:prop_large_y} the behaviour of $\Delta^g(x = \xi, \mu)$ for the LO PDF set MMHT2014~\cite{Harland-Lang:2014zoa} as an input at $\mu_0$ = 1 GeV. On the left, $\Delta^g$ is represented with an uncertainty at initial scale produced by the function $g$ with parameter $\alpha = 1$. As compared to the case on the right where $\alpha = 2$, this corresponds to significantly smaller uncertainties for $z > \xi$ . Yet, the results are similar, proving that the determining source of uncertainty comes from the region around $z = \xi$.

\begin{figure}
    \centering
    \includegraphics[scale=0.62]{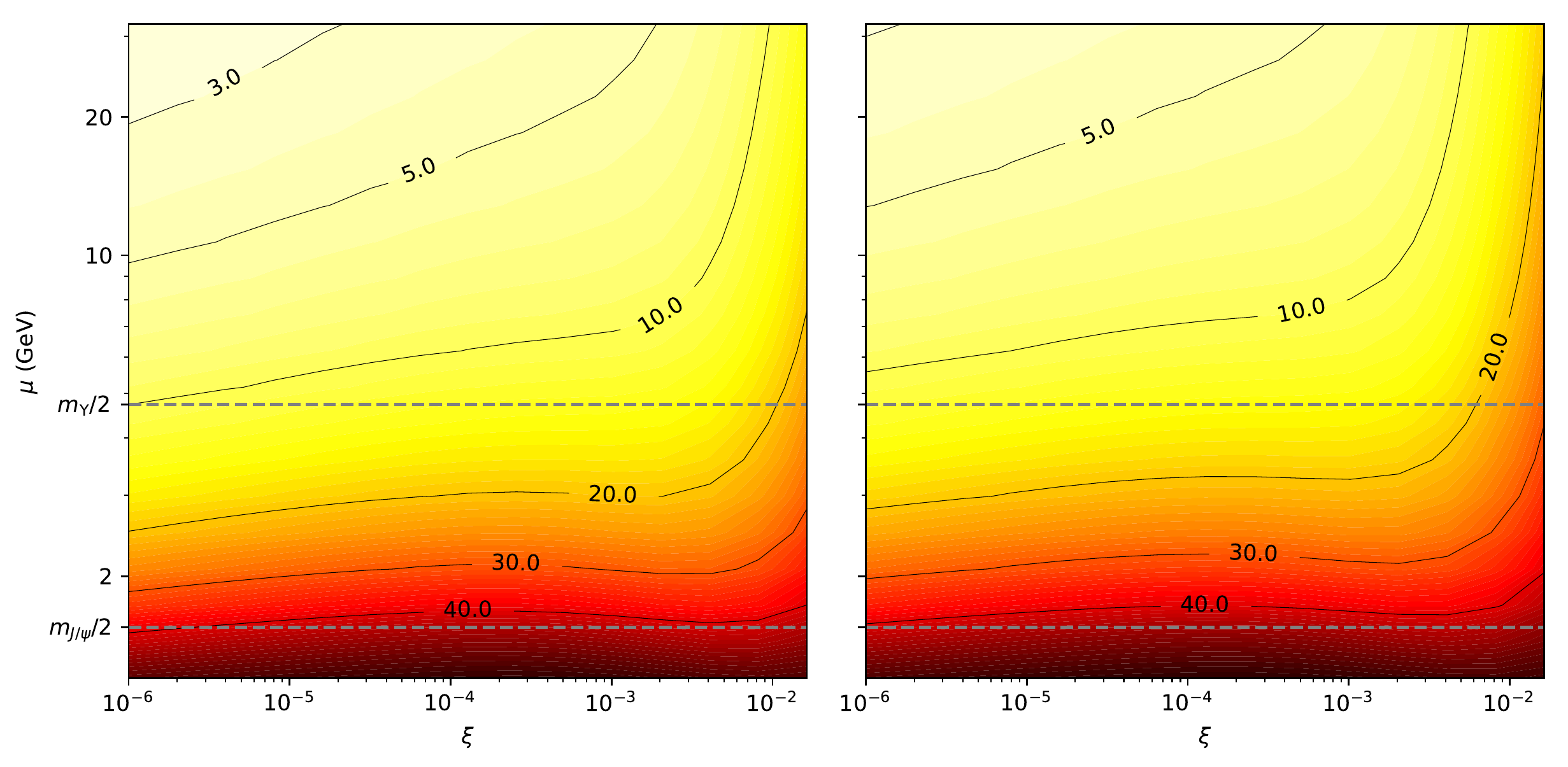}
    \caption{Behaviour of $\Delta^q(x = \xi, \xi, \mu)$ in \% as a function of $\xi$ and $\mu$ for the LO PDF set MMHT2014 used as input at scale $\mu_0 = 1$~GeV. On the left, we use the uncertainty profile with $\alpha = 1$ while on the right with $\alpha = 2$ (see Fig.~\ref{fig:unc} and Eq.~\eqref{eq:unc_profile}).}
    \label{fig:prop_large_y_sea}
\end{figure}

As expected from Eq.~\eqref{eq:fhoeizo}, when $\mu$ is very small, the uncertainty is close to $g(1) = 60\%$. At the scale $\mu=1.5$~GeV, corresponding to $J/\psi$ production, the uncertainty of the $\xi$ reconstruction of the gluon GPD can be conservatively estimated to be 10 to 20\% for a large range of values of $\xi$. At the scale $\mu=4.7$~GeV corresponding to $\Upsilon$ production, the uncertainty on the gluon GPD drops approximately to 2 to 4\%. Although at LO HVMP is only sensitive to the gluon GPD, we show the result for the light quark sea GPD in Fig.~\ref{fig:prop_large_y_sea}. There the situation is much direr, with an uncertainty of order 40\% for $J/\psi$ production, and 10\% for $\Upsilon$. This is easily understood by looking at the evolution operators shown on Fig.~\ref{fig:operators}. The contributions of the large-$z$ region for the $qq$ operator is significantly less than that of the $gg$ operator, meaning that the effect of washing away of uncertainty by evolution is much less effective for quarks than for gluons.

The behaviour of $\Delta^g$ as a function of $\xi$ results from two opposite trends. When $\xi$ is still rather large, the decrease of $\xi$ triggers a very quick increase of the contribution of the region $z \gg \xi$, resulting in a quick reduction of the uncertainty. However, when $\xi$ decreases below roughly $10^{-3}$, $\Delta^g$ stabilises. The cause of this behavior is an interplay between the increase of the operator weights $\Gamma^{ab}$ in the large-$z$ region and the steep increase of the PDFs themselves at small-$z$. As a consequence, the power-law behavior of the PDFs at small-$z$ stabilises the dominance of the large-$z$ region in the evolution. Rather than $\xi$, it is $\mu$ which plays the most important role in the quality of the approximation.

We believe that one can use the value of $\Delta^a(x, \xi, \mu)$ as a conservative estimate of the uncertainty generated by the extrapolation to vanishing skewness of GPDs. As mentioned in the previous section, the extrapolation to $\xi \rightarrow 0$ is not only relevant to relate GPDs to PDFs in the context of small-$x$ HVMP, but also to the extraction of impact-parameter distributions (IPDs) \eqref{eq:ipds}, a major aspect of the hadron tomography programme.

The uncertainty in the extrapolation to vanishing skewness depends on three aspects: the choice of the low-lying scale $\mu_0$, the choice of the uncertainty profile $g$ \eqref{eq:unc_profile}, and the knowledge of the PDFs at the initial scale $\mu_0$. Let us review them one by one. One would find advantage in taking $\mu_0$ as low as possible to increase the leverage of evolution, which we have highlighted in Fig.~\ref{fig:prop_large_y} to be a dominant factor in the reduction of $\Delta^a$. However, low scales are associated with large missing higher-order corrections that we do not take into account here. This limitation would be mitigated by the use of higher-order evolution. Indeed, our proposal can be used without any change with any perturbative order of GPD evolution. One could thus envision to use a modified evolution kernel resumming large $\log(x)$ in the spirit of various attempts to modify the DGLAP equation in the small-$x$ region (see for instance Ref.~\cite{Ball:2017otu} and references therein).

That $\Delta^a$ depends on the choice of uncertainty profile \eqref{eq:unc_profile} is the consequence of the fact that we are trying to quantify something that is fundamentally unknown, namely the $\xi$-dependence of GPDs at the low-lying scale $\mu_0$. Fig.~\ref{fig:prop_large_y} highlights that a dominant factor in the uncertainty profile is $g(1)$, which characterises the difference between $F^a(x= \xi,\mu_0)$ and $f^a(x, \mu_0)$. In this study, we have used $g(1) = 60\%$ based on a general argument of large skewness ratios encountered in the literature, but it unfortunately boils down to the physicist to make an assumption on the bound of "reasonable" uncertainties on the GPDs at the initial scale. It is obvious that an arbitrarily large uncertainty at initial scale $\mu_0$ would result in an arbitrarily large uncertainty at a hard scale as well.

Finally, $\Delta^a$ depends on the PDF profile at $\mu_0$. It means that our uncertainty estimate is Bayesian in nature, in the sense that the uncertainty in relating the GPDs to the PDFs at the hard scale $\mu$ depends on our prior knowledge of the PDFs at the scale $\mu_0$. This can be understood physically: if the PDFs at the low scale $\mu_0$ increase only moderately at small-$x$, then the dominance of the large-$z$ region is easier to establish, and the uncertainty $\Delta^a$ is smaller. The uncertainty $\Delta^a$ should therefore be evaluated with respect to our current best knowledge of PDFs at small-$x$ and low scales. To fully leverage the Bayesian viewpoint on the uncertainty of the vanishing $\xi$ extrapolation, one could consider the following strategy:
\begin{enumerate}
    \item Start from the current best knowledge of PDFs at the low scale $\mu_0$.
    \item Evolve this PDF set using $\xi$ dependent GPD evolution to the hard scale of the process. This produces GPD-like objects with a $\xi$-dependence entirely resulting from evolution. 
    \item Compute the cross section of the process using these GPD-like objects. Evaluate the compatibility of this prediction with the actual measurement, taking into account that the prediction contains, in addition to the uncertainty inherited from the PDFs at the scale $\mu_0$, an additional source of uncertainty represented by $\Delta^a$.
    \item Update the PDF knowledge through a reweighting strategy to evaluate the impact of the actual measurement on the knowledge of the PDFs, while taking into account the experimental uncertainty of the measurement, the theoretical uncertainty of the initial PDF knowledge, and the systematic uncertainty of the vanishing $\xi$ extrapolation.
\end{enumerate}
Reweighting strategies as described in the last step are a well-established technique for PDFs, and are increasingly used for higher-dimensional parton distributions (see for instance Refs.~\cite{Ball:2010gb, Ball:2011gg, Dutrieux:2021ehx}).

The proposal we have developed in this section is similar in its general strategy to the one based on the Shuvaev transformed which was already applied for instance in Refs.~\cite{Jones:2013pga, Flett:2019pux, Flett:2021fvo}. We clarify in the next section how the two modelling strategies differ.

\section{Revisiting the modelling of GPDs through the Shuvaev transform}

The Shuvaev transform~\cite{Shuvaev:1999fm,Shuvaev:1999ce} relates the representation of GPDs in momentum space as we presented them so far to the representation of GPDs in the space of conformal moments. Conformal moments are defined as~\cite{Diehl:2003ny}
\begin{equation}
    \mathcal{O}_n^a(\xi, t, \mu) = \frac{\Gamma(n+1-p_a) \Gamma(p_a+3/2)}{2^{n-p_a} \Gamma(n+3/2)}\,\xi^{n-p_a} \int_{-1}^1 \mathrm{d}x\,C_{n-p_a}^{(p_a+3/2)}\left(\frac{x}{\xi}\right) F^a(x, \xi, t, \mu)\,,
\end{equation}
where $C_n^{(p_a+3/2)}$ are Gegenbauer polynomials and $\Gamma$ is the Euler gamma function generalising factorials. The prefactor with $\Gamma$ functions is introduced such that in the limit $t = 0$ and $\xi = 0$, the conformal moments are exactly equal to the Mellin moments of the PDF, defined by
\begin{equation}
    \mathcal{M}^a_n(\mu) = \int_{-1}^1 \mathrm{d}x\,x^n\,f^a(x, \mu)\,.
\end{equation}
Conformal moments are particularly suitable to solve the LO GPD evolution equations~\cite{Efremov:1979qk, Makeenko:1980bh}. For instance, for $n$ even -- that corresponds to the quark non-singlet GPD -- the $\xi$ and $\mu$ dependence is factorised as follows
\begin{equation}
    \mathcal{O}_n^q(\xi, t, \mu) = \mathcal{O}_n^q(\xi, t, \mu_0) \left(\frac{\alpha_s(\mu)}{\alpha_s(\mu_0)}\right)^{\gamma_n / 2\beta_0}\,,
    \label{eq:loevol}
\end{equation}
where the anomalous dimensions $\gamma_n$ are the same as those that govern the DGLAP evolution of PDFs. In the singlet case, only conformal moments associated to the same odd value of $n$ mix between the gluon and quark singlet GPDs. The Shuvaev operator, which we denote as $\mathcal{S}^a(x, \xi; n)$, allows one to relate the two representations of GPDs through
\begin{equation}
    F^a(x, \xi, t, \mu) = \mathcal{S}^a(x, \xi, n) \ast \mathcal{O}_n^a(\xi, t, \mu)\,,
    \label{eq:reconstruct}
\end{equation}
where we use the $\ast$ notation to indicate the action of the Shuvaev operator on an analytical continuation of the conformal moments.

Defining the $\xi$-dependence of GPDs can equivalently take the form of defining the $\xi$-dependence of conformal moments. The modelling proposal of GPDs at small-$\xi$ based on the Shuvaev transform consists in approximating the conformal moments in the limit $\xi \ll 1$ with their value at $\xi = 0$, \textit{i.e.} in defining a GPD model whose conformal moments are constant in $\xi$ and equal to
\begin{equation}
    \mathcal{O}^a_n(\xi, t, \mu) \equiv \int_{-1}^1 \mathrm{d}x\,x^{n-p_a} F^a(x, \xi = 0, t, \mu)\,,\quad \xi\ll 1\,,
\end{equation}
which are exactly the Mellin moments of the PDF if $t = 0$. We will again set aside the question of the modelling of the $t$-dependence, which should ideally be the subject of a flexible parametrisation at the level of conformal moments. At $t=0$, this model greatly simplifies the expression of Eq.~\eqref{eq:reconstruct}, since it can be turned into the following form
\begin{equation}
    F^a(x, \xi, \mu) \equiv \mathcal{S}'^a\left(\frac{\xi}{x}, \frac{\textbf{x'}}{\xi}\right) \ast f^a(\textbf{x'}, \mu)\,,
    \label{eq:shuvaevproposal}
\end{equation}
where $\mathcal{S}'^a$ is now the composition of the Shuvaev operator $\mathcal{S}^a$ and the Mellin transform (see Appendix~\ref{appendixShuvaev} for details). We use the boldface character $\textbf{x'}$ so that no ambiguity arises on the actual integration variable subtended by the symbol $\ast$. An equivalent formalism related to the assumption of $\xi$-independence of conformal moments has been discussed in Ref.~\cite{Musatov:1999xp}. The proposal has been extended in Ref.~\cite{Noritzsch:2000pr} which provides a technical fix so that the reconstructed GPDs do not extend outside the physical support $x \in [-1, 1]$.

\begin{figure}
    \centering
    \includegraphics[scale=0.7]{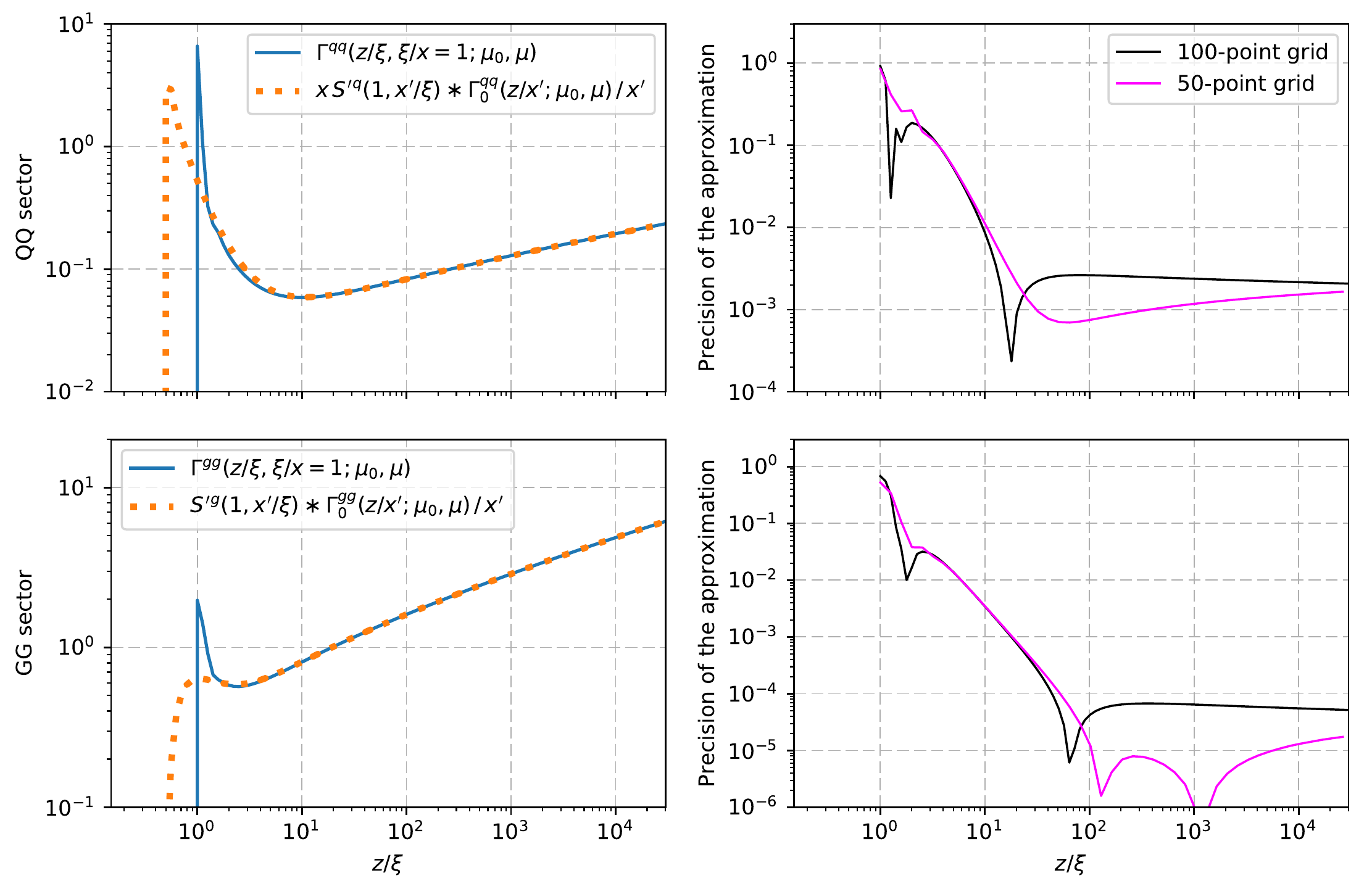}
    \caption{Quality of the approximation of LO evolution operators $\Gamma^{ab}(z / \xi, \xi / x; \mu_0, \mu)$ by Eq.~\eqref{eq:testshuvaev}, as a function of $z / \xi$ in the case where $\xi = x$, $\mu_0$ = 1~GeV and $\mu$ = 4.7~GeV. The $qq$ sector is displayed in upper-left panel while the $gg$ sector in the lower-left panel. On the right, the relative precision of the approximation is given for two different discretisations of the evolution operators, either with 50 or 100 points. The dips in the curves correspond to a change of sign in the difference between the true value of the LO evolution operator and its approximation by Eq.~\eqref{eq:testshuvaev}.}
    \label{fig:shuvaev_test}
\end{figure}

Let us show how this modelling ($\xi$-independent conformal moments) is deeply related to the one we presented in the previous section (GPD = PDF at a low scale $\mu_0$ and $\xi$-dependent evolution applied to reach $\mu$). First we note the following remarkable property linking the DGLAP LO evolution operators $\Gamma^{ab}_0$ introduced in Eq.~\eqref{eq:oppdf} to the  general LO evolution operators defined in Eq.~\eqref{eq:opdef}: provided $z \gg \xi$, and given any $\mu > \mu_0$,
\begin{equation}
    \frac{1}{x}\Gamma^{ab}\left(\frac{z}{\xi}, \frac{\xi}{x}; \mu_0, \mu\right) \approx \frac{1}{x^{p_a}} \mathcal{S}'^a\left(\frac{\xi}{x}, \frac{\textbf{x'}}{\xi}\right) \ast \frac{1}{\textbf{x'}}\Gamma_0^{ab}\left(\frac{z}{\textbf{x'}}; \mu_0, \mu\right)\,.
    \label{eq:testshuvaev}
\end{equation}
In other words, in the limit where $z \gg \xi$, the $\xi$-dependence of the general LO evolution operator can be entirely reproduced by the $\xi$-dependence of $S'$. Fig.~\ref{fig:shuvaev_test} shows the excellent quality of this approximation in the $qq$ and $gg$ sectors for $\xi = x$, where we expect the largest deviations between the DGLAP evolution operators and the GPD ones. In the region $z \gg \xi$, we observe a numerical agreement of the order of $10^{-3}$ for quarks and $10^{-4}$ for gluons. Using various discretisations of the evolution operators with the evolution code shows that numerical uncertainties dominate the discrepancy at large-$z$. On the contrary, for $z \lesssim 10 \xi$, both discretisations give generally similar results: in this region, the approximation of Eq.~\eqref{eq:testshuvaev} gives significant discrepancies compared to the true value of the GPD evolution operator. It is obvious from the expression of $S'$ in Appendix~\ref{appendixShuvaev} that Eq.~\eqref{eq:testshuvaev} cannot hold for small values of $z$: as can also be observed on Fig.~\ref{fig:shuvaev_test}, whereas the GPD evolution operator should be 0 for $z < \xi$ in the case where $x = \xi$, the r.h.s. of Eq.~\eqref{eq:testshuvaev} gives non-vanishing contributions down to $z = \xi/2$.

Under the assumption that the contribution of the region $z \lesssim 10\xi$, where the approximation of Eq.~\eqref{eq:testshuvaev} is not very accurate, is negligible in the evolution of the GPD from $\mu_0$ to $\mu$, it is possible to write the right-hand side of~\eqref{eq:shuvaevproposal} as
\begin{align}
    \mathcal{S}'^a\left(\frac{\xi}{x}, \frac{\textbf{x'}}{\xi}\right) \ast f^a(\textbf{x'}, \mu) &= \mathcal{S}'^a\left(\frac{\xi}{x}, \frac{\textbf{x'}}{\xi}\right) \ast \sum_{b=q,g,...} \int_0^1 \frac{\mathrm{d}z}{\textbf{x'}}\, \Gamma_0^{ab}\left(\frac{z}{\textbf{x'}}; \mu_0, \mu\right) \, f^b(z, \mu_0) \,,\\
    &\approx x^{p_a} \sum_{b=q,g,...} \int_0^1 \frac{\mathrm{d}z}{x}\,\Gamma^{ab}\left(y, \frac{\xi}{x}; \mu_0, \mu\right)\, f^b(z, \mu_0) \,,
    \label{eq:trans}
\end{align}
where we used Eq.~\eqref{eq:oppdf} in the first line, and applied the approximation of Eq.~\eqref{eq:testshuvaev} in the second. Since we assumed dominance of the $z \gtrsim 10\xi$ region in the integral over $z$, we can also assume that the skewness between incoming and outgoing four-momenta $z + \xi$ and $z - \xi$ is negligible, leading to the approximation that $f^b(z, \mu_0) \approx z^{-p_b} F^b(z, \xi, \mu_0)$. Hence
\begin{align}
    \mathcal{S}'^a\left(\frac{\xi}{x}, \frac{\textbf{x'}}{\xi}\right) \ast f^a(\textbf{x'}, \mu) &\approx x^{p_a}\sum_{b=q,g,...} \int_0^1 \frac{\mathrm{d}z}{x}\,\Gamma^{ab}\left(y, \frac{\xi}{x}; \mu_0, \mu\right)\, \frac{F^b(z, \xi, \mu_0)}{z^{p_b}}= F^a(x, \xi, \mu)\,.
    \label{eq:reshuvaevproposal}
\end{align}
Therefore, the modelling strategy expressed in Eq.~\eqref{eq:shuvaevproposal} relies on the same argument as the one presented before, that is that there exists a lower scale $\mu_0$ such that the large-$z$ region dominates the evolution. However, any reference to $\mu_0$ disappears in the final formulation of the proposal because the relation between general LO and DGLAP evolution operators of Eq.~\eqref{eq:testshuvaev} holds independently of $\mu$. However, the disappearance of $\mu_0$ in the modelling strategy with the Shuvaev transform deprives us of a crucial tool to evaluate the uncertainty of the procedure.

Quantification of the systematic uncertainty of the Shuvaev procedure has been so far mostly treated with generic estimates, summarised as $\mathcal{O}(\xi^2)$ at LO and $\mathcal{O}(\xi)$ at NLO, based on considerations on the polynomial expansion of Gegenbauer moments of GPDs as functions of $\xi$~\cite{Shuvaev:1999ce}. The strict mathematical validity of these estimates has been criticised for instance in Ref.~\cite{Diehl:2007zu, Kumericki:2009ji} where it is argued that, in general, truncating the $\xi$ expansion of conformal moments -- in this case to its first term -- and performing their analytical continuation to apply the Shuvaev transform are non-commutative operations. Therefore, when the PDFs exhibit specific singularities, the reconstructed GPD would not produce the correct forward limit. Even assuming that the generic uncertainty estimates of $\mathcal{O}(\xi^2)$ at LO and $\mathcal{O}(\xi)$ at NLO are mathematically founded for most of the phenomenologically relevant parametrisations of PDFs, as argued in Ref.~\cite{Martin:2008gqx}, it remains to clarify what is the actually size of these terms, which should at least depend on the hard scale $\mu$. On the contrary, the uncertainty we have presented in the previous section does not scale as $\mathcal{O}(\xi^2)$ or $\mathcal{O}(\xi)$, and it is rather steady of the order of several percents to tens of percents at very small values of $\xi$.

Finally, the modelling with the Shuvaev transform is based on the property of LO evolution of conformal moments. It means that only the specific $\xi$-dependence introduced by the LO evolution operator can be precisely reproduced, whereas our proposal can be straightforwardly extended to higher orders.

\section{Conclusion}

We propose a method to quantify the uncertainty in relating GPDs to PDFs at small-$x_B$. We observe that this uncertainty decreases quickly when the hard scale of the process increases, but presents a mild dependence on the value of $\xi \approx x_B/2$, contrary to the typical estimates $\mathcal{O}(\xi^2)$ or $\mathcal{O}(\xi)$ provided in the literature. We are able to give conservative estimates of the uncertainty for heavy-vector-meson production of $J/\psi$ of the order of 10 to 20\% if the hard scale 1.5~GeV is used~\cite{Jones:2015nna, Jones:2016ldq}, and of 2 to 4\% for $\Upsilon$ at a hard scale of 4.7~GeV. We show that our method relies on the same general arguments as the one based on the Shuvaev transform.

We finally remark that deriving the $\xi$-dependence of GPDs from PDFs at small-$\xi$ effectively subtracts a degree of freedom to the modelling of GPDs in this kinematic domain. This entails a natural solution to the \textit{deconvolution problem} of factorised observables. It has been explicitly demonstrated in Ref.~\cite{Bertone:2021yyz} that for deeply-virtual Compton scattering, a process closely related to HVMP, it is not possible to perform an extraction of GPDs from experimental data in a model independent way at next-to-leading order due to the poor conditioning of the problem. The crux of the issue is essentially that the dimension of the space of kinematic variables available to experiments, $(x_B, t, Q^2)$, is smaller than the space of variables of GPDs,  $(x, \xi, t, \mu^2)$. This problem is considerably tamed by the model we propose in this article. The uncertainty on the extrapolation to vanishing skewness which we have derived represents then a measure of the systematic uncertainty associated to this choice of regularisation for the deconvolution problem at small-$\xi$.

\appendix

\section{Formulas for the Shuvaev transform} \label{appendixShuvaev}

The operator $S'^a$ resulting from the composition of the Shuvaev operator $S^a$ and the  Mellin transform acts on a PDF $f^a(w)$ in the following way~\cite{Shuvaev:1999ce}
\begin{equation}
    \begin{array}{c}
    \displaystyle \mathcal{S}'^a\left(\frac{\xi}{x}, \frac{\textbf{x'}}{\xi}\right) \ast f^a(\textbf{x'}) =\\
    \displaystyle \int_{-1}^1 \mathrm{d}x' \left[\frac{2}{\pi}\ \mathrm{Im} \int_0^1 \mathrm{d}s \left(\frac{4s(1-s)}{(x+\xi(1-2s))^{1+p_a}}\sqrt{1-\frac{4sx'(1-s)}{x+\xi(1-2s)}}\right)^{-1}\right]\frac{\mathrm{d}}{\mathrm{d}x'}\left(\frac{f^a(x')}{|x'|}\right)\,,
    \end{array}
    \label{eq:zhduoipza}
\end{equation}
where $p_a = 1$ for gluons ($a=g$) and 0 for quarks ($a = q$). A study of the support of the integral yields that, for $x > \xi > 0$, Eq.~\eqref{eq:zhduoipza} can be expressed as 
\begin{align}
    \mathcal{S}'^a\left(\frac{\xi}{x}, \frac{\textbf{x'}}{\xi}\right) \ast f^a(\textbf{x'}) &= \int_{x/2+\sqrt{x^2-\xi^2}/2}^1 \mathrm{d}x'  \,C^a\left(\frac{\xi}{x}, \frac{x'}{\xi}\right)\,\frac{\mathrm{d}}{\mathrm{d}x'}\left(\frac{f^a(x')}{x'}\right)\,, \label{eq:finalmethod}\\
    C^a\left(\frac{\xi}{x}, \frac{x'}{\xi}\right) &= \frac{2}{\pi}\ \mathrm{Im} \int_{s_1}^{s_2} \mathrm{d}s \left(\frac{4s(1-s)}{(x+\xi(1-2s))^{1+p_\alpha}}\sqrt{1-\frac{4sx'(1-s)}{x+\xi(1-2s)}}\right)^{-1}\,,
\end{align}
\begin{equation}
    s_{1,2} = \frac{1}{2}+\frac{\xi}{4x'}\mp\frac{\sqrt{4x'^2+\xi^2-4xx'}}{4x'}\,. 
\end{equation}
It is clear that the integral is finite if $x > \xi$ as we never get to integrate up to the problematic boundaries $s = 0$ and 1. For $x = \xi$,
\begin{equation}
    C^a\left(1, \frac{x'}{x}\right) = \frac{2x^{1+p_a}}{\pi}\ \mathrm{Im} \int_{x/2x'}^{1} \mathrm{d}s \left(\frac{s}{2^{p_a-1}(1-s)^{p_a}}\sqrt{1-\frac{2sx'}{x}}\right)^{-1}\,.
\end{equation}
Considering that $p_a = 0$ for quarks and $1$ for gluons, we find
\begin{equation}
    C^q\left(1, \frac{x'}{x}\right) = \frac{x}{\pi}\ \mathrm{Im} \int_{x/2x'}^{1} \frac{\mathrm{d}s}{s\sqrt{1-\frac{2sx'}{x}}} = -\frac{2x}{\pi}\ \textrm{arctan}\left(\sqrt{\frac{2x'}{x}-1}\right)\,.
\end{equation}
\begin{equation}
    C^g\left(1, \frac{x'}{x}\right) = \frac{2x^2}{\pi}\ \mathrm{Im} \int_{x/2x'}^{1} \mathrm{d}s\,\frac{1-s}{s\sqrt{1-\frac{2sx'}{x}}} = 2x C^q\left(1, \frac{x'}{x}\right) + 2 \frac{x^3}{x' \pi}\sqrt{\frac{2x'}{x}-1}\,.
\end{equation}

\newenvironment{acknowledgement}{\relax}{\relax}
\begin{acknowledgement}
\section*{Acknowledgements}

This work is supported in part in the framework of the GLUODYNAMICS project funded by the "P2IO LabEx (ANR-10-LABX-0038)" in the framework "Investissements d’Avenir" (ANR-11-IDEX-0003-01)
managed by the Agence Nationale de la Recherche (ANR), France.

\end{acknowledgement}

%%%%%%%% Bibliography 
\bibliographystyle{utphys}   % Remember we use title in the biblio
\bibliography{bibliography}

\end{document}